\documentclass[12pt]{emulateapj}

\newcommand{\omcp}{\hbox{$\omega$ Cen}}

\renewcommand{\deg}{\mbox{$^{\circ}$}}

\def\Min{${}^{\prime}$\llap{.}}
\def\Sec{${}^{\prime\prime}$\llap{.}}
\def\min{${}^{\prime}$}
\def\sec{${}^{\prime\prime}$}

\def\ltsim{\, {}^<_\sim \,}

\shorttitle{On the absolute age of the globular cluster NGC~3201} 
\shortauthors{Bono et al.}

\begin{document}
%
\title{On a new Near-Infrared method to estimate the absolute ages  
of star clusters: NGC~3201 as a first test case\altaffilmark{1}}

\author{
G.\ Bono\altaffilmark{2,3},
P.\ B.\ Stetson\altaffilmark{4},
D.\ A.\ VandenBerg\altaffilmark{5},
A.\ Calamida\altaffilmark{6},
M.\ Dall'Ora\altaffilmark{7},
G.\ Iannicola\altaffilmark{3},
P.\ Amico\altaffilmark{6}, 
A.\ Di Cecco\altaffilmark{2},
E.\ Marchetti\altaffilmark{6},
M.\ Monelli\altaffilmark{8},
N.\ Sanna\altaffilmark{2},
A.\ R. \ Walker\altaffilmark{9}
M.\ Zoccali\altaffilmark{10}
R.\ Buonanno\altaffilmark{2},
F.\ Caputo\altaffilmark{3},
C.\ E.\ Corsi\altaffilmark{3},
S.\ Degl'Innocenti\altaffilmark{11,12}
S.\ D'Odorico\altaffilmark{6}
I.\ Ferraro\altaffilmark{3},
R.\ Gilmozzi\altaffilmark{6},
J.\ Melnick\altaffilmark{6},
M.\ Nonino\altaffilmark{13}, 
S.\ Ortolani\altaffilmark{14},
A.\ M.\ Piersimoni\altaffilmark{15},
P.\ G.\ Prada Moroni\altaffilmark{11,12},
L.\ Pulone\altaffilmark{3}, 
M.\ Romaniello\altaffilmark{6}, and   
J.\ Storm\altaffilmark{16}   
}

\altaffiltext{1}{Based on near infrared observations made with ESO telescopes 
SOFI@NTT, La Silla; MAD@VLT Paranal, projects: 66.D-0557, 074.D-0655, ID96406 
and with the CTIO telescope ISPI@4m Blanco, La Serena. 
Based on optical data collected with ESO telescopes and retrieved from the 
ESO Science Archive Facility.\\   
This research used the facilities of the Canadian Astronomy Data Centre 
operated by the National Research Council of Canada with the support of 
the Canadian Space Agency.}
\altaffiltext{2}{Dept. of Physics, UniToV, via della Ricerca Scientifica 1, 00133 Rome, Italy; bono@roma2.infn.it}
\altaffiltext{3}{INAF--OAR, via Frascati 33, Monte Porzio Catone, Rome, Italy}
\altaffiltext{4}{DAO--HIA, NRC, 5071 West Saanich Road, Victoria, BC V9E 2E7, Canada} 
\altaffiltext{5}{Dept. of Physics and Astronomy, UVIC, Victoria, BC V8W 3P6, Canada} 
\altaffiltext{6}{ESO, Karl-Schwarzschild-Str. 2, 85748 Garching bei Munchen, Germany}
\altaffiltext{7}{INAF--OACN, via Moiariello 16, 80131 Napoli, Italy}
\altaffiltext{8}{IAC, Calle Via Lactea, E38200 La Laguna, Tenerife, Spain}
\altaffiltext{9}{CTIO--NOAO, Casilla 603, La Serena, Chile}
\altaffiltext{10}{PUC, Dept. Astronomia y Astrofisica, Casilla 306, Santiago 22, Chile}
\altaffiltext{11}{Dept. of Physics, Univ. Pisa, Largo B. Pontecorvo 2, 56127 Pisa, Italy}
\altaffiltext{12}{INFN, Sez. Pisa, via E. Fermi 2, 56127 Pisa, Italy}
\altaffiltext{13}{INAF--OAT, via G.B. Tiepolo 11, 40131 Trieste, Italy}
\altaffiltext{14}{Dept. of Astronomy, Univ. of Padova, Vicolo dell' Osservatorio 5, 35122 Padova, Italy}
\altaffiltext{15}{INAF--OACTe, via M. Maggini, 64100 Teramo, Italy}
\altaffiltext{16}{AIP, An der Sternwarte 16, D-14482 Potsdam, Germany}

\date{\centering drafted \today\ / Received / Accepted }

\begin{abstract}
We present a new method to estimate the absolute ages of stellar systems. 
This method is based on the difference in magnitude between the main 
sequence turn-off (MSTO) and a well defined knee located along the lower 
main sequence (MSK). This feature is caused by the collisionally induced 
absorption of molecular hydrogen and it can be easily identified in 
near-infrared (NIR) and in optical-NIR color-magnitude diagrams of stellar 
systems. We took advantage of deep and accurate NIR images collected with 
the Multi-Conjugate Adaptive Optics Demonstrator temporarily available 
on the Very Large Telescope and of optical images collected with 
the Advanced Camera for Surveys Wide Field Camera on the Hubble Space 
Telescope and with ground-based 
telescopes to estimate the absolute age of the globular NGC~3201 using both 
the MSTO and the $\Delta$(MSTO-MSK). We have adopted a new set of cluster 
isochrones and we found that the absolute ages based on the two methods agree 
to within one sigma. However, the errors of the ages based on the $\Delta$(MSTO-MSK) 
method are potentially more than a factor of two smaller, since they are 
not affected by uncertainties in cluster distance or reddening.    
Current isochrones appear to predict slightly bluer ($\approx$0.05 mag) 
NIR and optical-NIR colors than observed for magnitudes fainter than the MSK.
\end{abstract}


\maketitle

\section{Introduction}
The absolute ages of Galactic Globular Clusters (GGCs) is a crossroad 
of several astrophysical problems (VandenBerg et al.\ 1996; Chaboyer 1998; 
Castellani 1999). This parameter provides: {\em i)}~a lower limit to the 
age of the universe (Buonanno et al.\ 1998; Stetson et al.\ 1999; 
Gratton et al.\ 2003); {\em ii)}~robust constraints on the physics 
adopted in stellar evolutionary models (Salaris \& Weiss 1998; 
Cassisi et al.\ 1999; VandenBerg et al.\ 2008; Dotter et al.\ 2008), 
and {\em iii)}~the chronology for the assembly of the halo, bulge and 
disk of the Milky Way (Carraro et al. 1999; Rosenberg et al.\ 1999; 
Zoccali et al.\ 2003; de Angeli et al.\ 2005).
However, estimates of absolute GC ages are still hampered by
uncertainties in the distance, extinction and the reddening law (Renzini 
1991; Bono et al.\ 2008), the overall
metallicity scale (Rutledge et al.\ 1997; Kraft \& Ivans 2003), as
well as the
detailed distribution of heavy-element abundances (Asplund et al.\ 2005), 
and photometric zero-points (Stetson 2000). 

Proposed ways of overcoming some of these thorny problems include the
use either of a different clock (white dwarf cooling sequence, 
Richer et al.\ 2006), or different photometric systems (Str\"omgren 
bands, Grundahl et al.\ 1998) or different diagnostics (luminosity 
function, Zoccali et al.\ 2000; Richer et al.\ 2008). However, 
these alternative approaches to estimate the absolute 
ages of GCs can be impeded by limiting magnitude, 
by photometric accuracy, or by sample completeness.  
Simultaneous and self-consistent interpretation of optical and NIR photometry 
is another way to refine current distance and reddening estimates.  
In fact, as shown in this study (see also Calamida et al.\ 2009) 
deep $J,K$ photometry reveals the existence of a well-defined ``knee''
in the lower main sequence. The position of this feature is, at fixed chemical
composition, essentially independent of the cluster age. Therefore, the difference
either in magnitude or in color with the turn-off can provide robust estimates of
the absolute age.

Consider the GC NGC~3201, which has several interesting properties. It is
relatively nearby and it suffers from appreciable foreground reddening 
($\mu=13.32\pm0.06$, $E(B-V)=0.30\pm0.03$, 
Piersimoni et al.\ 2002; $\mu=13.36\pm0.06$, $E(B-V)=0.25\pm0.02$, 
Layden \& Sarajedini 2003; Mazur et al.\ 2003). Accurate abundances estimates 
exist for both iron ([Fe/H]=$-1.54\pm0.10$, Kraft \& Ivans 2003; 
Covey et al.\ 2003) and $\alpha$-elements ([$\alpha$/Fe]$\sim$ 0.2--0.4, 
Pritzl et al. 2005). 
However, NGC~3201 is affected by field contamination and by differential 
reddening (Piersimoni et al.\ 2002; Kravtov et al.\ 2009). Owing to these 
problems, we still lack an accurate estimate of its absolute age 
(Layden \& Sarajedini 2003).

\section{Observations and data reduction}

Near-Infrared (NIR) data ($J,K_s$) were collected in two observing 
runs with different pointings of the NIR camera SOFI (Field of View [FoV]
$\sim$ 5\min$\times$ 5\min; scale 0\Sec29 /px)
on the New Technology Telescope (NTT; ESO, La Silla). We obtained
132 $J$-band images with individual exposure times from 3 
to 6\,s and median seeing 0\Sec8. We also collected 161 $K_s$-band 
images with exposure times from 10 to 12\,s and median seeing 
0\Sec6. The total exposures per band are $\approx 10$  
($J$) and $\approx 40$ minutes ($K_s$). These images unevenly cover an
area of $\approx$ 20\min $\times$ 18\min across the cluster 
center (see Fig.~1). NIR data ($J,K_s$) were also collected with the 
NIR camera ISPI (FoV $\sim$10\Min5$\times$10\Min5; scale 0\Sec30 /px)
on the 4m Blanco telescope (CTIO; La Serena). We collected 15
$J$-band images, each of which was the co-added result of six individual
5\,s exposures; these had median seeing 0\Sec74 arcsec. We also obtained
15 $K_s$-band images, consisting of eight co-added images with individual
exposure times of 5\,s, and median seeing 0\Sec75 arcsec. The total
exposures per band are thus 450\,s in $J$ and 600\,s in $K_s$.  

These data were supplemented with deep NIR data ($J,K_s$) 
from the Multi-Conjugate Adaptive Optics Demonstrator 
(MAD) temporarily available on the Very Large Telescope 
(VLT; ESO, Paranal). MAD is a prototype instrument performing 
real-time correction for atmospheric turbulence (Marchetti et al.\ 2006).
Its infrared camera employs a 2048$\times$2048 Hawaii-II infrared detector
with a scale of 0\Sec028 /px for a total FoV of $\approx$ 1\min.
During the first Science Demonstration of MAD four 1\min$\times$1\min\ fields 
were observed in a region located in the S-W corner of NGC~3201.
Five guide stars with visual magnitudes ranging from 
11.8 to 12.9 distributed on a circle of 2 arcmin diameter 
concentric to the field were used. For each 
pointing we collected three $J$-band and five $K_s$-band images of 240 sec 
(DIT=10, NDIT=24). The seeing during the observations changed from 
0.6 to 0.9 arcsec ($J$-band) and from 0.8 to 1.3 arcsec ($K_s$-band).
The full width at half-maximum (FWHM) measured on the images ranges from 
0.07 to 0.10 arcsec. 
We ended up with a catalog containing $\sim$54,100 
stars with at least one measurement in each of two different bands.
The top panel of Fig.~1 shows the spatial distribution of the different 
NIR datasets\footnote{Following Harris (1996) we assumed for the center 
of the cluster the coordinates: $\alpha=10^h17^m$36\sec8,  
$\delta$=--46\deg24\min40\sec}. 

The $V$ and $I$ data used in this investigation come from the
database of original and archival observations that have been collected,
reduced, and calibrated by one of the authors (PBS) in an ongoing effort to 
provide homogeneous photometry on the Landolt (1992) photometric system 
for a significant fraction of GGCs (Stetson 2000).
For NGC~3201, we currently have a catalog consisting of $\sim$180,000 stars 
with at least two measurements in each of $B$, $V$, and $I$; 
these span an area with extreme dimensions on the sky of $\sim$40\min  
(E-W) by $\approx$50\min (N-S).  These data were obtained in the course
of 12 ground-based observing runs with various telescopes, as well as with 
the Wide Field Planetary Camera 2 (WFPC2) and the Advanced Camera for Surveys 
Wide Field Camera (ACS-WFC) on the Hubble Space Telescope (HST).
The accuracy of the zero-points of the optical catalog is believed to be
better than 0.01$\,$mag (see the bottom panel of Fig.~1).

Initial photometry on individual images was performed with {\tt DAOPHOT IV},
followed by simultaneous photometry for the 359 NIR and the 558 ground-based
optical images with {\tt ALLFRAME} (Stetson, 1994). The HST images were reduced
in a separate ALLFRAME run.  The instrumental IR magnitudes were transformed
into the 2MASS photometric system. The standard error for one observation of 
one star ranges from $\sim 0.03$ to $\sim 0.06\,$mag, depending upon the 
filter ($J$,$K$) and the night.  However, with $\sim$22,000 individual
observations of more than 1,000 2MASS stars in $J$ and $\sim$26,000 observations
in $K$, we expect that our results should be on the 2MASS photometric system to 
$\ltsim 0.02\,$ mag in the two bands.

\section{Comparison between theory and observations}

The NIR CMDs present the obvious advantage of being minimally affected by 
uncertain and possibly differential reddening. However, image 
quality, angular resolution, and photometric precision have 
partially hampered the use of NIR photometry in crowded fields. Data 
plotted in the top panel of Fig.~2 show the {\em K,J-K} CMD based on 
data collected with NIR cameras on 4m telescopes (black dots, 
SOFI@NTT, ISPI@4m Blanco) and with MAD on VLT (red dots). 
The error bars plotted in the right side show that the photometric 
precision is of the order of a few hundredths of a magnitude over most
of the magnitude range, increasing to $\sim 0.1$ mag at our adopted 
detection limit ($K\approx 20$, $J\approx 20.7$). 
Note, in particular, that the MS in the low-mass regime 
shows a well defined knee. This feature is mainly caused by the 
collisionally induced absorption of H$_{2}$ at NIR wavelengths 
(Saumon et al.\ 1994). According to theory, the color and the shape of 
the bending depends on the metal content. However, the magnitude and 
color of the bend are, at fixed chemical composition, essentially 
independent of cluster age. This feature offers the unique opportunity 
to anchor cluster isochrones, and in turn to estimate the absolute 
age either as a color or as a magnitude difference between the 
bend and the cluster TO. The knee is a robust prediction, 
since for (M$\approx$ 0.5-0.4 $M_\odot$) the stellar structures
are minimally 
affected by uncertainties in the treatment of convection, 
given that convective motions are nearly adiabatic (Saumon et al.\ 2008).  
In passing we note that a similar feature was already detected in 
two GCs (\omcp, Pulone et al.\ 1998; M4, Pulone et al.\ 1999) and 
in the Galactic bulge (Zoccali et al.\ 2000) using deep and high 
angular resolution NIR ($J$,$H$) data collected with NICMOS@HST.    
To estimate the ridge line we represented each star detected in the
$J,K$-bands by a Gaussian kernel with a sigma equal to the photometric
uncertainty. The individual Gaussians were summed and we identified the  
peaks along the MS. This ridge line was smoothed with a spline and 
the TO (red circle) and the MS knee (MSK, purple circle) were 
estimated as the points showing the minimum curvature in the two 
different color ranges (see also Table~1).  

To estimate the absolute age of NGC 3201, we have used the new set
of isochrones for $Y=0.250$, [Fe/H] $= -1.50$, and [$\alpha$/Fe]
$\approx 0.4$ from the VandenBerg et al.~(2010, in preparation) 
compilation.  [A constant
enhancement was not assumed for all $\alpha$-elements; rather, the
Cayrel et al.\ (2004) determinations of individual [element/Fe] ratios 
in metal-deficient stars were adopted.]  Fully consistent opacities 
and the latest nuclear reactions, along with treatments of H/He 
diffusion and turbulent mixing very close to those described by 
Proffitt \& VandenBerg (1991) and Proffitt \& Michaud (1991), 
respectively, were employed.  
Moreover, the free parameters in their simple treatment of turbulence 
were set so that the predicted variation in the surface helium abundance 
during the main-sequence and subgiant phases agreed well with that 
reported by Korn et al.\ (2007). The latter had notable success explaining 
the observed abundances of a number of the heavy elements, using models 
that allowed for the settling and radiative accelerations of the 
metals; consequently, their prediction of the dependence of helium 
abundance on evolutionary state represents the current best estimate. 
Although VandenBerg et al.\ have not treated the diffusion of the metals
their isochrones are very similar to those used in the Korn et al.\ 
investigation, both morphologically and insofar as the predicted age 
at a fixed turnoff luminosity are concerned.

The isochrones were transformed from the theoretical to the observed 
plane using the color-temperature (CT) relations derived by
Casagrande, VandenBerg, \& Stetson (2010, in preparation) from the 
latest MARCS model atmospheres (Gustafsson et al.\ 2008).  Small zero-point
adjustments (0.03 mag) were applied to colors of the isochrones so that the 
synthetic colors for stars fainter than the turnoff agreed well with those 
derived from the IRFM-based, empirical CT relations for dwarf stars by Casagrande
et al.~(2009).  Consistent fits of these isochrones to most of the CMDs
for NGC~3201 that can be derived from $BVRIJK_s$ photometry indicated a
clear preference for a true distance modulus $\mu_0 \sim 13.35\pm0.11$ and a
reddening $E(B-V) \approx 0.24\pm0.02$.

The extinction in the NIR bands was estimated using the McCall (2004) 
relations. The colored lines plotted in the top panel of Fig.~2 show 
the comparison with three isochrones at $t=$10,12 and 14 Gyr. However, 
the model isochrones in this CMD appear to be $\approx0.05$ mag bluer 
than observed for magnitudes fainter than the MSK. 

To overcome subtle uncertainties in the cluster reddening, we performed 
a comparison between theory and observations using
a Wesenheit $K$-band\footnote{$WK$=$K$-$A_K$/($A_J$-$A_K$) [$J$-$K$]
mag, where the $A_J$ and $A_K$ are the selective absorption according to 
McCall's relation.} magnitude. This magnitude is reddening free, 
so the reddening uncertainty in the $WK$,$J$-$K$ CMD occurs only in the
horizontal direction.  Data plotted in the bottom panel 
of Fig.~2 show even more clearly the systematic drift in color of 
cluster isochrones when compared with observations.     

To further constrain the possible culprit(s) for this discrepancy
we cross-correlated the NIR ($J,K$) photometry based on MAD 
images with the optical ($V,I$) photometry based on ACS images. 
The top panels of Fig.~3 show the $K$,$I-K$ (left) and the 
$K$,$V-K$ (right) optical-NIR CMD.   
Data plotted in these panels display several interesting features. 
{\em i)}-- The MS knee is still present, although less evident than 
in the NIR CMD. However, the approach we devised allows us to provide 
accurate magnitude and color estimates for the age indicators (see 
Table~1).  
{\em ii)}-- The same isochrones that in the NIR ($K$,$J-K$) 
CMD attain systematically bluer colors fainter than MSK provide---assuming 
identical values for distance and reddening---a reasonable fit to both evolved 
and MS stars. However, the isochrones again attain systematically bluer 
colors below the MS knee ($K\ge$18.5 mag). 
The same outcomes apply if the comparison is performed using the 
reddening free Wesenheit $K$-band magnitude and the same 
optical-NIR colors (see the bottom panels in Fig.~3).    
 
As a further test of the isochrones we performed the same 
comparison for an optical $I$,$V-I$ CMD. In particular,  
we took advantage of the deep and precise photometry from ACS 
and ground-based images. Data in Fig.~4 display a well 
defined cluster sequence ranging from the subgiant branch 
to the very low-mass regime ($I\sim24$, $V-I\sim 2$ mag). 
Note that a handful of candidate cluster white dwarfs shows up in the 
left bottom corner ($I\sim24.5$, $V-I\sim 0.25$ mag). The MS knee 
is less evident than in the optical-NIR CMDs, but both the high 
number of stars and the photometric precision permit a robust 
identification of the age indicators (see Table~1). 
Interestingly, we found that the same isochrones in 
the NIR and optical-NIR CMDs provide a good fit to the 
observations over the entire magnitude and color range. 
The isochrones are only slightly redder for magnitudes fainter 
than the MSTO. 
This clearly indicates that optical-NIR, and to a minor extent 
NIR, CMDs are an acid test of the precision of both evolution 
and atmosphere models. Even small changes in the input physics 
can be easily identified in the quoted CMDs.

The comparison between theory and observations highlights two relevant 
findings: {\em i)}-- the adopted chemical composition is appropriate for 
NGC3201, since cluster isochrones simultaneously account for both optical 
and optical-NIR CMDs; {\em ii)}-- the main reason in the discrepancy 
between theory and observations in the NIR and in the optical-NIR CMDs 
appears to be the $K$-band and to a small extent the $I$-band.
The isochrones attain colors that are slightly either bluer (K,J-K) or redder
(I,V-I) than the observed MSK. An independent and preliminary test performed
using the same isochrones, but a different set of CT relations (Casagrande et
al.\ 2010) show a smaller color discrepancy and very similar cluster ages.

\section{Results and discussion}

The adopted cluster isochrones for ages ranging from 10 to 14 Gyr  
indicate that the magnitude/age derivative of MSTO stars changes from 
$\Delta$$V$/$\Delta$t$\sim$ 0.093, to $\Delta$$I$/$\Delta$t$\sim$ 0.080 
and to $\Delta$$K$/$\Delta$ t $\sim$ 0.062 mag/Gyr. This means that 
the age sensitivity decreases by $\approx$35\% when moving from the 
$V$ to the $K$-band. On the other hand, the color/age derivative of MSTO
stars is $\Delta$$(V-I)$/$\Delta$t$\sim$ 0.013 for optical bands,  
$\Delta$$(V-K)$/$\Delta$t$\sim$ 0.032, $\Delta$$(I-K)$/$\Delta$t$\sim$ 0.019
for optical-NIR bands and $\Delta$$(J-K)$/$\Delta$t$\sim$ 0.010 mag/Gyr for 
NIR bands. These color-age derivatives indicate that optical ($V-I$) 
and NIR ($J-K$) colors attain similar sensitivities, while the optical-NIR 
($V-K$, $I-K$) colors are, as expected, a factor of two-three more 
sensitive.      
A decrease of 0.2 dex in iron content implies a $\sim$20\% decrease  
in magnitude/age derivatives, while an increase of the same amount 
causes minimal changes ($\le$5\%). The same changes in chemical 
composition have minimal effects on the color/age derivatives.      

However, the reddening law by McCall (2004) gives selective 
absorptions of $A_I/A_V$$\sim$0.538\footnote{To account for the difference 
in the effective wavelength between the I-band adopted by McCall 
($\lambda_{eff}^I$=7977 \AA) and by Landolt ($\lambda_{eff}^I$=8250 \AA)  
the selective absorption $A_I$/$E(B-V)$ was changed, following 
Cardelli et al.\ (1989),  from 1.714 to 1.652.}, $A_K/A_V$$\sim$0.112 mag 
and  reddening corrections of $E(V-K)/E(V-I)$$\sim$ 1.923, 
$E(I-K)/E(V-I)$$\sim$ 0.923 and $E(J-K)/E(V-I)$$\sim$ 0.343 mag. 
These numbers indicate that the $K$ magnitudes and $J-K$  
colors are respectively one order of magnitude and more than a factor 
of two less affected by reddening uncertainties ($\sim$10\%) than 
the $V-I$ colors and the $I$ magnitudes.  

To constrain the error budget in the absolute age estimates of GCs, 
based on the MSTO, we also need to account for uncertainties 
affecting the distance determination. The uncertainty in the GC distance 
depends on the adopted 
standard candle and is still of the order of 5\% (Bono et al.\ 2008).     
Using the quoted error budget we estimated the absolute magnitude 
of the MSTO in the optical and in NIR bands (see Table~1). The error 
budget on absolute magnitudes is, as expected, dominated by the 
uncertainties in cluster distances. It is on average $\sim$0.11 mag 
for $M_K$, $M_I$ and for the Wesenheit magnitudes.  
The uncertainty of the chemical composition and the cluster isochrones
will be discussed in a future paper (Dall'Ora et al.\ 2010, in 
preparation). Using the quoted isochrones 
we found absolute age estimates and uncertainties (see Table~1) for NGC~3201 
that agree, within one sigma, with similar estimates in the literature. 
The weighted mean of the three age estimates based on the $K$-band is 
$11.48\pm1.27$, while that based on the $WK$-band is $11.55\pm1.53$ Gyr.  
Optical and NIR (see Table~1) age estimates agree within one sigma.  
The relative ages (De Angeli et al.\ 2005) are typically affected by 
smaller errors, but they rely on predictions of evolved evolutionary 
phases (HB, RGB) affected by different theoretical uncertainties. 

We also estimated the absolute ages of NGC~3201 from the difference 
in magnitude between the MSTO and the MSK using the same approach
and the same isochrones. Values listed in Table~1 indicate that ages
based on the new method agree, within the errors, with those based on 
the MSTO. The weighted mean of the three age estimates based on the $K$-band 
is $10.66\pm0.58$ and $11.63\pm0.58$ Gyr based on the $I$-band, but the errors 
are more than a factor of two smaller. The same applies on average to the 
new age estimates based on the $WK$ ($10.82\pm0.68$ Gyr) and on the 
$WI$-band.   
The new age estimates, indeed, do not include uncertainties in the distance 
and reddening; their uncertainties are dominated by intrinsic photometric 
errors ($\sigma_{J,K}\sim$0.02, $\sigma_{V,I}\sim$0.01--0.02 mag).

The new method to estimate the absolute
age of GCs requires accurate NIR photometry $\approx$2--3 
magnitude fainter than the TO region ($M_J\sim M_K\approx$4--6). 
This means apparent magnitudes $J\sim K\approx$19--21 mag 
for a good fraction of GGCs and $J\sim K\approx$14--17 mag 
for Galactic open clusters. The former limiting magnitudes 
can be reached with 8-10m class telescopes, while the latter 
are within the limits of 2-4m class telescopes. 
However, the Extremely Large Telescopes and JWST will allow us 
to apply this method to all the stellar systems in the Local 
Group---provided that suitable, local faint calibrating 
sequences exist.  

\acknowledgments
We thank C. Brasseur, L. Casagrande and A. Dotter for useful 
discussions concerning NIR CMDs. 
This project was supported by Monte dei Paschi di Siena 
(P.I.: S. Degl'Innocenti), PRIN-INAF2006 (P.I.: M. Bellazzini), 
PRIN-MIUR2007 (P.I.: G. Piotto).
MZ acknowledges Fondap Center for Astrophysics (15010003), CATA PFB-06
and  Fondecyt Regular \#1085278.
We are extremely grateful to ESO for the farsightedness in making 
available a valuable instrument (MAD) to perform accurate NIR photometry.
 

\begin{deluxetable}{lccccccrr}
\scriptsize
\tablewidth{0pt}
\tablecaption{Magnitudes and colors of Main Sequence Turn-Off 
(MSTO) and Main Sequence Knee (MSK) in different optical and NIR CMDs.}
\tablehead{
\colhead{CMD}&
\colhead{$m_{MSTO}$\tablenotemark{a}} &
\colhead{$CI_{MSTO}$\tablenotemark{b}}&
\colhead{$m_{MSK}$\tablenotemark{a}}  &
\colhead{$CI_{MSK}$\tablenotemark{b}} & 
\colhead{$M_{MSTO}$\tablenotemark{c}} & 
\colhead{$M_{MSK}$\tablenotemark{c}} & 
\colhead{$t_{MSTO}$\tablenotemark{d}} & 
\colhead{$t_{MSTO-MSK}$\tablenotemark{e}}  
}
\startdata
 $K$,~$J$-$K$& 16.26 & 0.41 & 18.30 & 0.77 &  2.83 & 4.87  & $11.8\pm2.3$ & $10.5\pm1.0$ \\ 
$WK$,~$J$-$K$& 15.95 & 0.41 & 17.77 & 0.78 &  2.59 & 4.41  & $12.0\pm2.8$ & $10.4\pm1.1$ \\ 

 $K$,~$I$-$K$& 16.26 & 1.06 & 18.31 & 1.73 &  2.83 & 4.88  & $11.8\pm2.3$ & $10.5\pm1.0$ \\ 
$WK$,~$I$-$K$& 15.91 & 1.06 & 17.72 & 1.63 &  2.56 & 4.37  & $11.3\pm2.6$ & $10.9\pm1.2$ \\

 $K$,~$V$-$K$& 16.22 & 1.96 & 18.34 & 3.10 &  2.78 & 4.90  & $11.0\pm2.1$ & $11.0\pm1.0$ \\
$WK$,~$V$-$K$& 15.92 & 1.96 & 17.81 & 3.14 &  2.57 & 4.46  & $11.4\pm2.6$ & $11.3\pm1.2$ \\

 $I$,~$V$-$I$& 17.28 & 0.91 & 20.46 & 1.46 &  3.53 & 6.71  & $11.5\pm2.0$ & $11.6\pm0.6$ \\
$WI$,~$V$-$I$& 16.22 & 0.91 & 18.71 & 1.43 &  2.87 & 5.36  & $11.6\pm2.1$ & $11.2\pm1.2$ \\
\enddata
\tablenotetext{a}{Apparent magnitudes of MSTO and MSK.}
\tablenotetext{b}{Color indices of MSTO and MSK.}
\tablenotetext{c}{Absolute magnitudes of MSTO and MSK.}
\tablenotetext{d}{Cluster age estimates (Gyrs) based on the absolute magnitude 
of MSTO. The uncertainties account for errors on photometry, reddening and 
distance.}
\tablenotetext{e}{Cluster age estimates (Gyrs) based on the difference in 
magnitude between MSTO and MSK. The uncertainties account for errors on 
photometry.}
\end{deluxetable}

\begin{figure}[!ht]
\begin{center}
\label{fig1}
\includegraphics[height=0.50\textheight,width=0.50\textwidth]{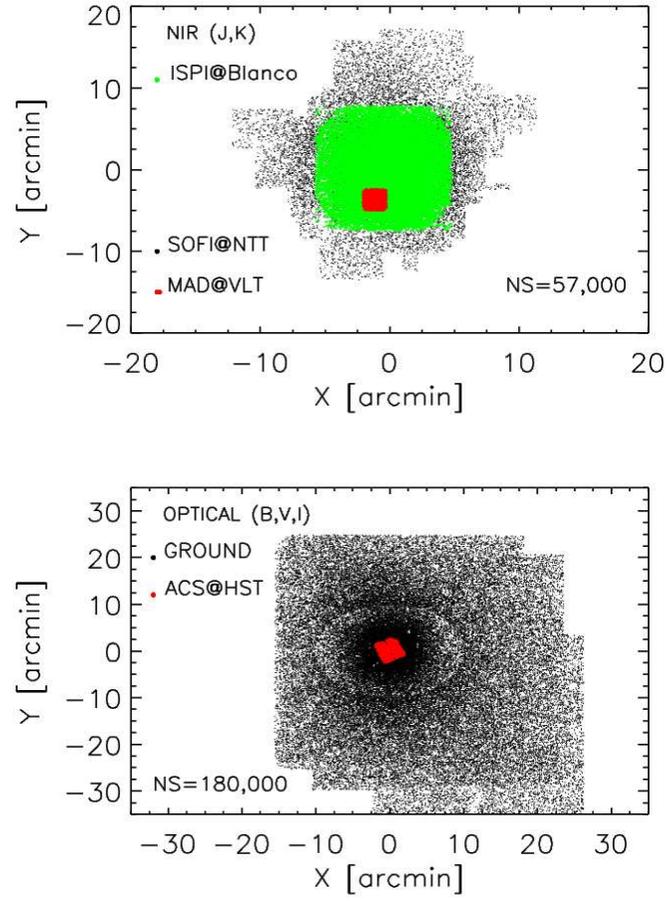}
\caption{Top--Radial distribution of stars observed in the NIR 
($J$,$K$) bands with different ground-based telescopes.   
Bottom--Same as the top, but for stars observed in the optical 
($V$,$I$) bands with ground-based telescopes and with HST. 
}
\end{center}
\end{figure}

\begin{figure}[!ht]
\begin{center}
\label{fig2}
\includegraphics[height=0.50\textheight,width=0.5\textwidth]{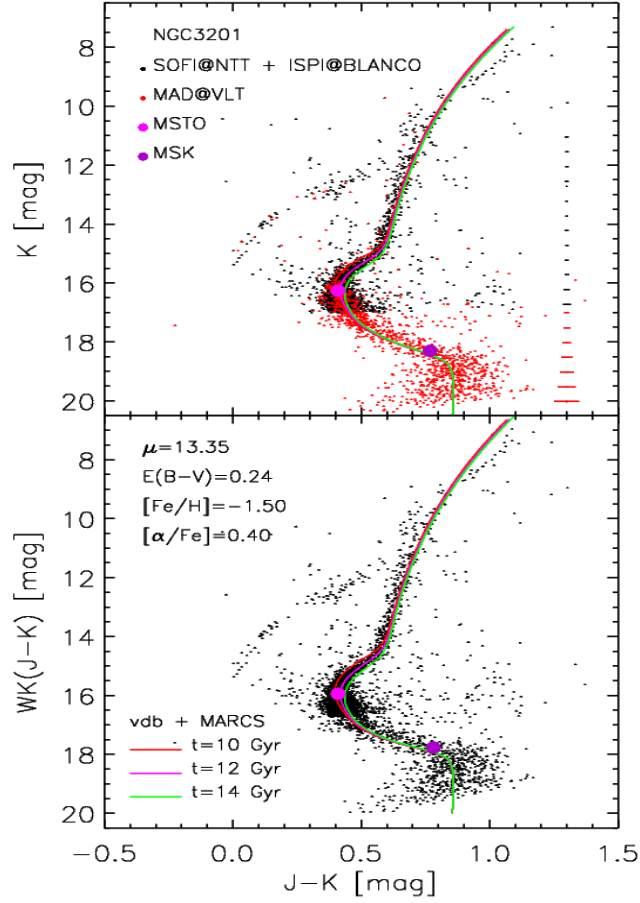}
\vspace*{1.05truecm}
\caption{Top -- NIR ($K$, $J-K$) CMD of NGC3201. Black and red dots display 
stars observed with SOFI@NTT, ISPI@Blanco and with MAD@VLT. The colored 
solid lines display a set of cluster isochrones by vandenBerg et al.\ (2009), 
at fixed chemical composition, and different ages (see labeled values). 
These isochrones were transformed into the observational plane using 
the CT relations based primarily on MARCS model atmospheres (see the
text). The adopted true distance and reddening are labeled. The large pink
and purple circles mark the position of MS Turn-Off and of the MS knee. 
Bottom -- Same as the top, but using the reddening free $WK$ magnitude. 
}
\end{center}
\end{figure}

\begin{figure}[!ht]
\begin{center}
\label{fig4}
\includegraphics[height=0.35\textheight,width=0.40\textwidth]{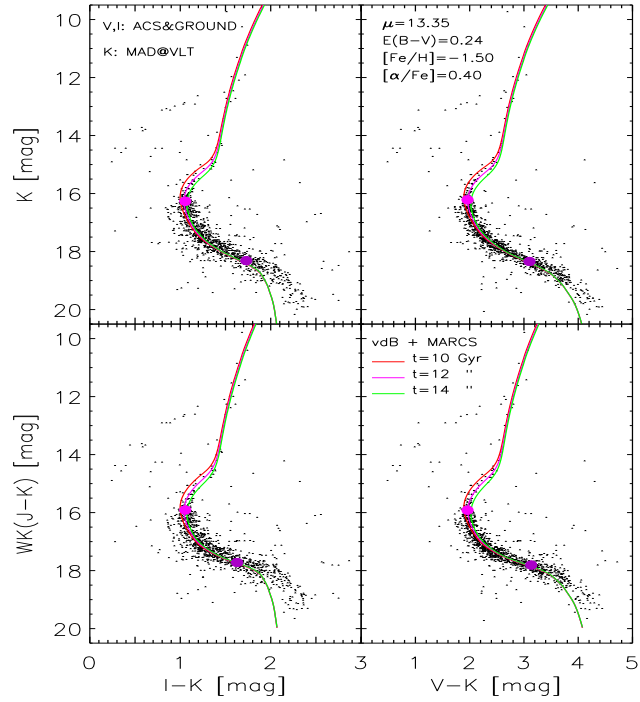}
\vspace*{0.71truecm}
\caption{Top -- $K$, $I-K$ (left) and $K$, $V-K$ (right) optical-NIR CMDs 
of NGC3201. Optical data were collected with ACS-WFC@HST, while the 
$K$-band data were collected with MAD@VLT. Symbols and lines are the 
same as in Fig.~2. Bottom -- Same as the top, but using the reddening 
free $W_K$ magnitude. 
}
\end{center}
\end{figure}

\begin{figure}[!ht]
\begin{center}
\label{fig4}
\includegraphics[height=0.35\textheight,width=0.40\textwidth]{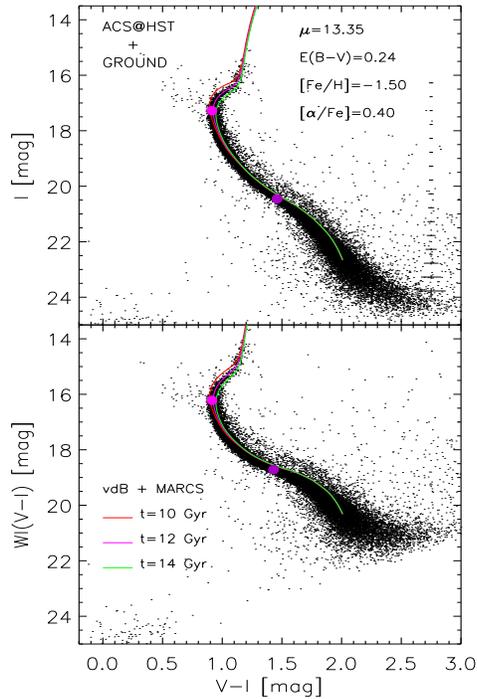}
\vspace*{0.71truecm}
\caption{Top -- Optical $I$, $V-I$ CMD of NGC3201. Data were collected with 
ACS-WFC@HST and ground-based telescopes. Symbols and lines are the same as 
in Fig.~2. The error bars on the right display intrinsic errors in magnitude 
and color. Bottom -- Same as the top, but using the reddening free $W_I$ 
magnitude. 
}
\end{center}
\end{figure}

\end{document}